\documentclass{optica-article}

\journal{optcon}

\articletype{Research Article}

\usepackage{lineno}

\begin{document}

\title{Spinning Metasurface Stack for Spectro-polarimetric Thermal Imaging}

\author{Xueji Wang,\authormark{1} Ziyi Yang,\authormark{1} Fanglin Bao,\authormark{1},Tyler Sentz,\authormark{1} and Zubin Jacob\authormark{1,*}}

\address{\authormark{1}Elmore Family School of Electrical and Computer Engineering, Birck Nanotechnology Center, Purdue University, West Lafayette, IN 47907\\}

\email{\authormark{*}zjacob@purdue.edu} 



\begin{abstract}
Spectro-polarimetric imaging in the long-wave infrared (LWIR) region plays a crucial role in applications from night vision and machine perception to trace gas sensing and thermography.  However, the current generation of spectro-polarimetric LWIR imagers suffer from limitations in size, spectral resolution and field of view (FOV). While meta-optics-based strategies for spectro-polarimetric imaging have been explored in the visible spectrum, their potential for thermal imaging remains largely unexplored. In this work, we introduce a novel approach for spectro-polarimetric decomposition by combining large-area stacked meta-optical devices with advanced computational imaging algorithms.  The
co-design of a stack of spinning dispersive metasurfaces along with compressive sensing and dictionary learning algorithms allows simultaneous spectral and polarimetric resolution without the need for bulky filter wheels or interferometers. Our spinning-metasurface-based spectro-polarimetric stack is compact ($<$ 10 x 10 x 10 cm), robust, and offers a wide field of view (20.5°). We show that the spectral resolving power of our system substantially enhances performance in machine learning tasks such as material classification, a challenge for conventional panchromatic thermal cameras.  Our approach represents a significant advance in the field of thermal imaging for a wide range of applications including heat-assisted detection and ranging (HADAR).
\end{abstract}

\section{Introduction}
The demand for high-resolution, information-rich image data has been amplified by the widespread industry adoption of machine learning algorithms\cite{Appl_Bio, Spectral_3D, Spectral_Raman2, Spectral_Raman, PhysicsInformed, tang2023active}. Meta-optics-enabled spectral and polarimetric imaging exhibits potential in meeting these rising data demands of learning algorithms within the visible spectrum\cite{MosaicSensor2, MosaicSensor3, PolImaging, huang2022full, Metasurface_Imaging, Metasurface_Imaging2, Metasurface_Imaging3, Metasurface_Imaging4, Metasurface_Imaging5}. Yet, the integration of meta-optics with infrared thermal imaging remains a relatively unexplored domain. Recent strides in heat-assisted detection and ranging (HADAR) have demonstrated the potential of thermal imaging\cite{HADAR} for machine perception tasks utilizing the infrared spectrum and the atmospheric transparency window. By combining spectral-resolved thermal imaging with artificial intelligence, HADAR offers a platform for machine perception through pitch darkness like broad daylight \cite{HADAR}. Here, our goal is to show that integrating meta-optics with infrared thermal imaging emerges as a crucial enabling factor for such next generation machine vision algorithms. Our work offers a pathway for realizing compact spectro-polarimetric thermal imagers beyond the conventional technologies which use push-broom, filter-wheel or interferometer modules. This new class of thermal imagers enhances the ability of machine learning algorithms for capturing information such as temperature, material composition, and surface morphology.

\begin{figure}[ht!]
\centering\includegraphics[width=13cm]{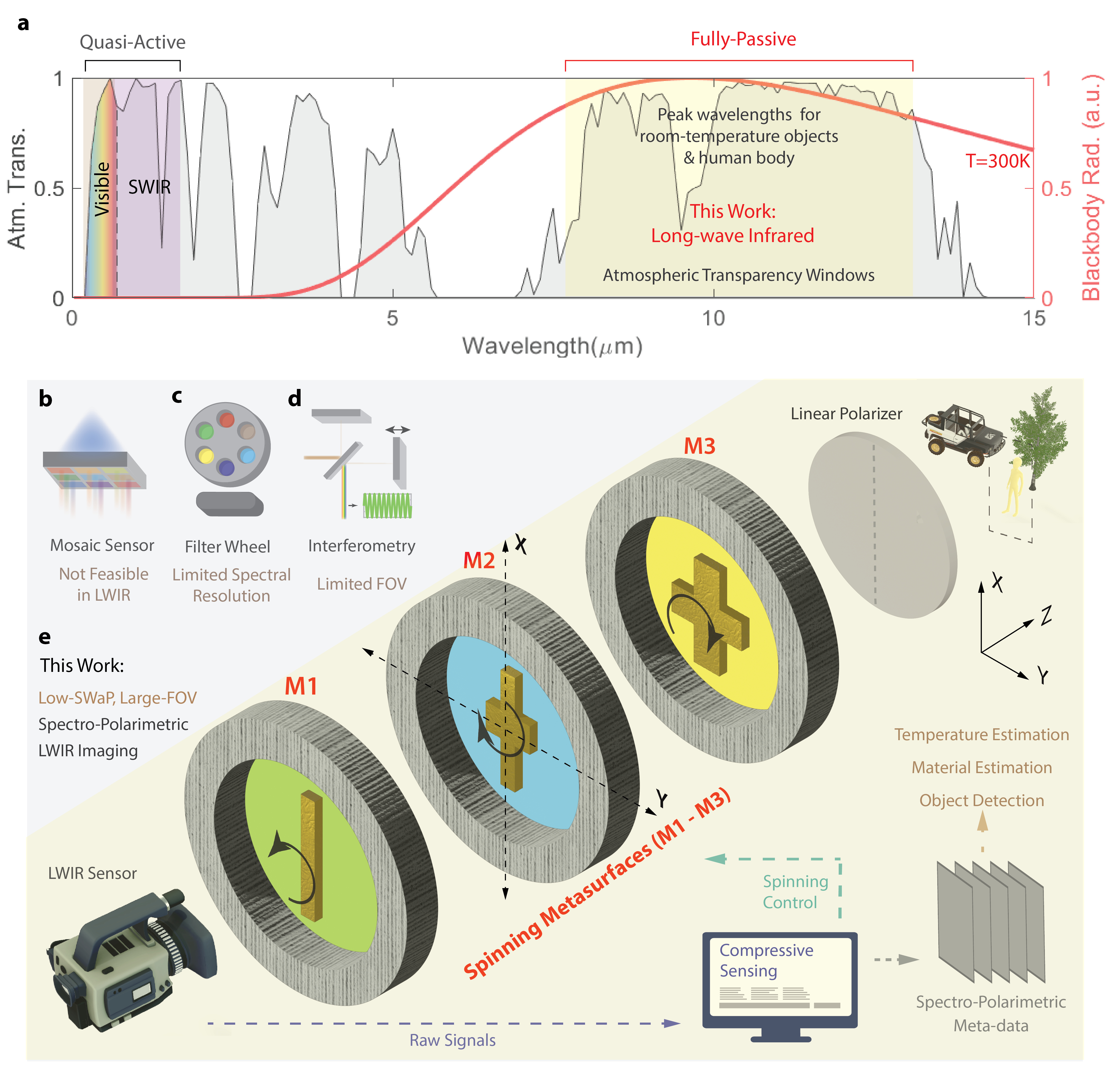}
\caption{Long-wave infrared (LWIR) spectro-polarimetric thermal imaging. (a) The room-temperature blackbody radiation (shown in red) and the atmospheric transmission spectrum (shown as a shaded area). The LWIR spectral region is crucial for thermal imaging due to its peaked room-temperature thermal radiation spectrum and the atmospheric transparency window. (b-d) Conventional methods for spectral imaging, such as using a mosaic sensor (b), a filter wheel (c), or interferometry (d), either pose limitations or are infeasible for LWIR thermal imaging. (e) In this study, we propose a new approach for spectro-polarimetric thermal imaging, achieved by combining large-area spinning metasurfaces and compressive sensing reconstruction algorithms}
\end{figure}

Within the broad thermal infrared spectrum, the long-wave infrared (LWIR) region stands out as especially advantageous for heat-assisted detection and ranging, as most objects at room temperature radiate thermal energy at these wavelengths (Fig.~1a). Moreover, the LWIR atmospheric transmission window facilitates the thermal radiation signal to propagate long distances\cite{RadCool_Fan}. However, the blurry nature of thermal images poses a significant challenge for machine vision algorithms. For example, perception tasks such as semantic segmentation and ranging rely on texture and contrast for learning features and show poor performance with panchromatic thermal imagery\cite{TI_Ghost}. HADAR overcomes the blurry nature of thermal images by decomposing the LWIR thermal radiation into its spectral channels\cite{HADAR}. Unfortunately, the commonly used mosaic filter technique for spectral decomposition (Fig.~1b)\cite{Spectral_Guo, NP_Spectral, MosaicSensor0, MosaicSensor1, MosaicSensor4, DiffuserCam} falls short for LWIR thermal imaging due to the restricted pixel count in LWIR focal plane arrays. Current HADAR spectral imagers rely on cumbersome filter wheels (Fig.~1c) with restricted spectral resolution, or fragile and bulky interferometers (Fig.~1d) with a confined field of view (FOV) \cite{HADAR}. Such constraints considerably hinder the widespread applicability and adoption of HADAR. 

To address these limitations, in this work, we harness the capabilities of large-area meta-optics in the infrared domain to create a platform for spectro-polarimetric thermal imaging. We achieve spectral decomposition by designing the dispersion and polarization through stacked metasurfaces, while the spectral reconstruction is accomplished using compressive sensing and dictionary learning algorithms. Our designs employ 2D structures with large feature sizes (> 1$\mu m$), which can be made using photo-lithography techniques, facilitating large-area fabrication and scalable manufacturing. Despite its compact form, our Spinning MetaCam delivers exceptional spectral resolution across diverse materials and can effectively unveil subtle polarimetric feature, enabling applications such as material classification . By leveraging meta-optics and advanced computational spectroscopy methods, we demonstrate three times higher accuracy for machine vision tasks than conventional panchromatic thermal imaging. In the future, the angular speed of the spinning stage can be increased to enable real-time video leading to widespread adoption of HADAR technology similar to  RADAR, LIDAR, and SONAR.

\section{Design of Spinning Metasurface Stack}

The architecture of our spinning-metasurface based spectro-polarimetric imaging system is depicted in Fig.~1e. It comprises of a broadband linear polarizer, three anisotropic and dispersive metasurfaces, and an LWIR imaging sensor. The polarizer is utilized to polarize the incoming thermal radiation signals, and the metasurfaces are utilized to realize spectral filtering. We design the metasurfaces with high anisotropy to produce distinct spectral responses for orthogonal polarizations. Additionally, the metasurfaces' dispersion rotates different wavelengths of radiation to varying polarization orientations. By using the metasurfaces in tandem and axially spinning the polarizer and metasurfaces to different angles, we obtain tunable transmission spectra that sample the incident thermal radiation in its spectral and polarimetric channels. We then reconstruct unknown spectra of imaging targets using compressive sensing and dictionary learning algorithms. Dictionary learning generates a set of basis functions that represent the unknown spectra in a sparse format \cite{DictLearning1}. Compressive sensing enables accurate reconstruction of the sparse spectra from limited number of measurements \cite{CompressedSensing}. Combining these two techniques enables accurate and stable spectral reconstruction in the presence of noise and measurement errors\cite{DictLearning2}. The four-dimensional spectro-polarimetric data generated by our system offers a wealth of physical information about an imaging target, making it a valuable tool for physics-driven machine vision\cite{PhysicsInformed, DeepSpectral}, facilitating various tasks such as object detection and semantic segmentation\cite{SpectralSegmantation, PolarSegmantation}.

\begin{figure}[ht!]
\centering\includegraphics[width=13cm]{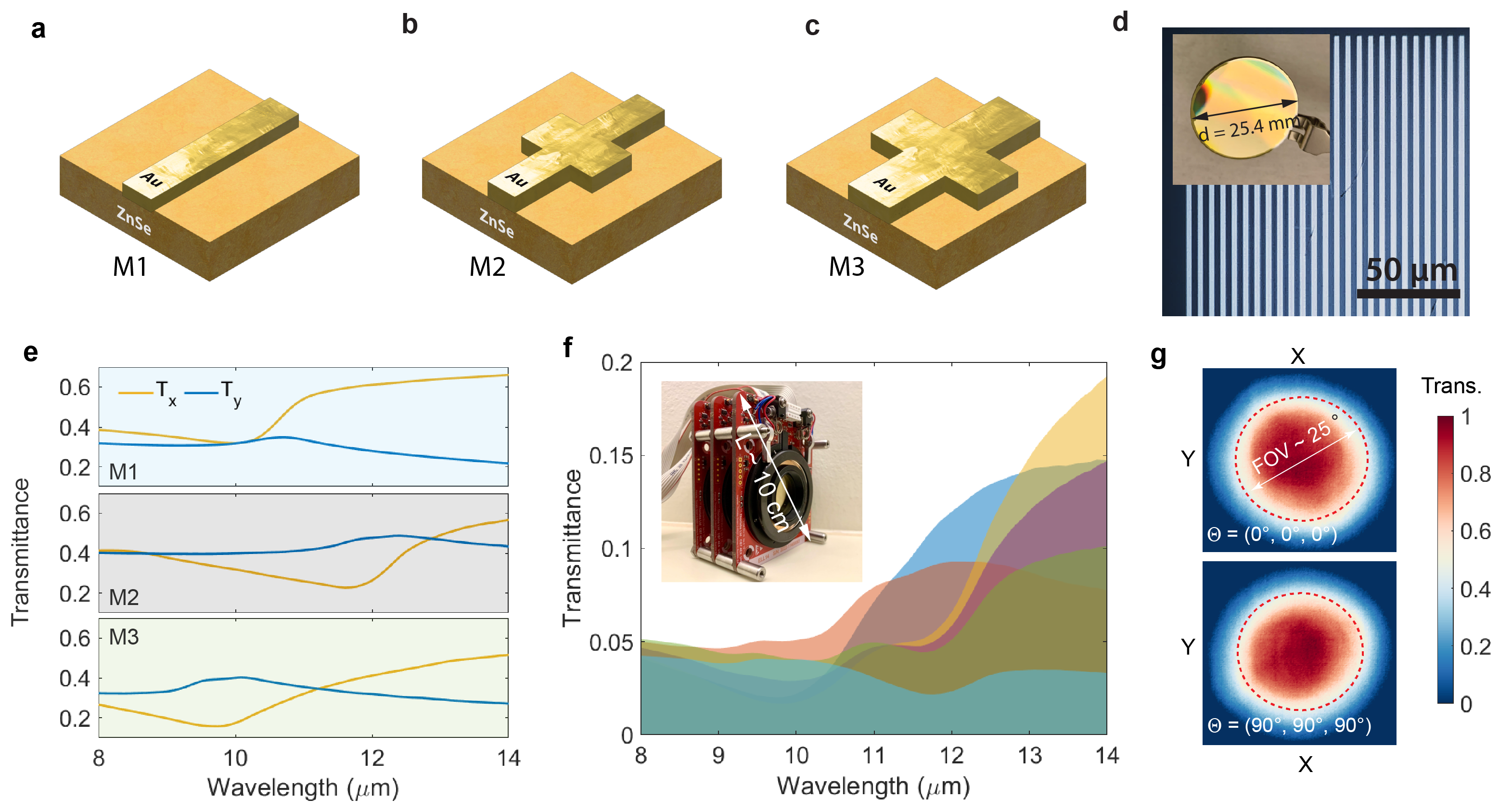}
\caption{Design and characterization of the spinning metasurfaces. (a-c) Schematics of the three different metasurface devices. (d) An optical microscope image of a fabricated metasurface. Inset: an optical image of a 1-inch-diameter device used for the imaging experiments, highlighting the large-area uniformity. (e) The measured polarized transmission spectra ($t_{ip}$ and $t_{is}$) of the three metasurfaces (M1 - M3), displaying strong anisotropy and distinctive dispersion. (f) The generated tunable transmission spectra by our spinning metasurface stack. The upper boundary of colored areas represents the transmission spectra corresponding to specific spinning angle combinations of the three metasurfaces. The colored areas highlight the differences between the generated transmission spectra. The low correlation between the generated transmission spectra is crucial for improving spectral reconstruction performance. Inset: an optical image of integrated spinning-metasurface module based on motorized rotatory mounts. The overall size of the module is smaller than 10 cm x 10 cm x 10 cm, making it a promising platform for next-generation high-contrast LWIR thermal imaging. (g) The normalized spatial transmittance of the module at two representative spinning angle combinations $\Theta = (0^{\circ},0^{\circ},0^{\circ})$ shown at the top  and $\Theta = (90^{\circ},90^{\circ},90^{\circ})$ illustrated at the bottom. The red circles correspond to a transmittance value of 0.5. Note that the spatial transmittance remains relatively consistent across all different $\Theta$ combinations, considering that the field of view is primarily limited by the diameters of the metasurface devices. The estimated field of view of the imaging system across the LWIR range is approximately 20.5$^{\circ}$.}
\end{figure}

To quantitatively describe the mechanism of the spinning-angle-controlled transmission spectra, we represent the spectro-polarimetric response of the metasurfaces using Jones matrices. Assuming that the transmission axis of the input linear polarizer is at 0 degrees
relative to x-axis, the Jones matrix $J_i$ of a metasurface i with the principal axis (p) at a spinning angle $\theta_i$ relative to the x-axis can be expressed as:

\begin{equation}
\begin{aligned} 
J_i(\theta_i, \lambda) & =R(-\theta_i)\cdot J_{M_i}(\lambda) \cdot R(\theta_i) 
\\ & =\left[\begin{array}{ll}cos(\theta_i) & -sin(\theta_i) \\ sin(\theta_i) & cos(\theta_i)\end{array}\right]\left[\begin{array}{cc}t_{ip}(\lambda)  & 0 \\ 0 & t_{is}(\lambda)\end{array}\right]\left[\begin{array}{ll}cos(\theta_i) & sin(\theta_i) \\ -sin(\theta_i) & cos(\theta_i)\end{array}\right]
\end{aligned} 
\end{equation}
where R is the rotation matrix, $J_{Mi}$ contains the anisotropic transmission of the metasurface $t_{ip}$ and $t_{is}$ along the two principle axes p and s. 

The Jones matrix of the three-metasurface assembly is given by:
\begin{equation}
J\left(\Theta, \lambda\right)=J_1\left(\theta_1, \lambda\right) \cdot J_2\left(\theta_2, \lambda\right) \cdot J_3\left(\theta_3, \lambda\right)\cdot
\label{eq:eq2}
\end{equation}

Thus, the total transmission spectrum of the three spinning metasurfaces strongly depends on the spinning-angle combinations $\Theta=(\theta_1$, $\theta_2$, $\theta_3)$ when the constituted metasurfaces are strongly anisotropic and dispersive, i.e. $t_{ip}(\lambda) \neq t_{is}(\lambda)$ (see Supplemental Document for the detailed analysis). We note that large differences between the spectral responses of the three metasurfaces (M1, M2, M3) are also introduced to minimize the correlations between the generated spectra, which can significantly improve the spectral reconstruction performance \cite{Yu_CompressedSensing}. We note that the metasurfaces do not necessitate rotation at fixed angular frequencies during the imaging process. As depicted in Eq. \ref{eq:eq2}, it is the static angular positioning of the metasurfaces that is pivotal for selectively filtering the incident thermal radiation in its spectral content. We emphasize that our design generates a large set of distinct transmission spectra with only three metasurfaces, while the total number of spectra in traditional mosaic array is limited to the number of metasurfaces/filters used \cite{Spectral_Guo, NP_Spectral, MosaicSensor0, MosaicSensor1, MosaicSensor2, MosaicSensor3, MosaicSensor4}.

Accordingly, we design the metasurfaces and experimentally achieve three key characteristics for optimized spectro-polarimetric imaging performance: 1) Strong anisotropy and dispersion for efficient wavelength decomposition; 2) High transmission and low self-emission for high signal to noise ratio (SNR); 3) Small angular dependence for a large FOV. The unit cell of the three designed metasurfaces are shown in Fig.~2 a-c. Strong dispersive anisotropy of the transmission spectra can be observed in Fig.~2e. Additionally, we emphasize that large-area devices are generally required for imaging applications to ensure sufficient numerical aperture. All the metasurfaces designed here have feature sizes larger than 1 $\mu m$. Large-area devices (25.4mm in diameter) with high structural quality and  uniformity (Fig.~2d) can be rapidly fabricated by standard photo-lithography techniques, enabling scalable manufacturing for practical applications. This is in strong contrast to recent works on miniaturized spectrometers\cite{Mini_Review, Mini2, Mini3, Mini1, Mini4, Mini5, Mini6, Mini7, Mini8}, where the device footprint is on the micrometer scale and thus not suitable for imaging applications.

The tunable transmission spectra produced by our spinning metasurfaces are shown in Fig.~2f. The distinct spectra are a result of the tuned spinning-angle combination $\Theta$. We integrate the three fabricated metasurfaces tandemly via compact rotatory mounts to independently control the rotation of each metasurface (Inset of Fig.~2f). We also optimize the spinning-angle combinations of the three spinning metasurfaces using genetic algorithms to generate largely uncorrelated transmission spectra for optimal spectral reconstruction performance (see Supplemental Document for details). Additionally, we note that increasing the number of metasurfaces can further improve the spectral resolution, but simultaneously reduces the SNR as the peak transmissions of the LWIR devices are limited to around 0.6 (see Supplemental Document figure S5 for details). However, our method has the potential to scale up into the hyperspectral regime by adding more high-transmission LWIR metasurfaces.

We also evaluate the FOV of our imaging module by integrating it with an LWIR thermal camera and capturing images of a large area uniform blackbody. To determine the spatial transmission efficiency, we normalize the signal counts of each pixel by the counts at the center of the images. We also define the angular range with transmittance above 0.5 as the effective FOV of a system. As seen in Fig.~2g, our spinning metasurface module has an FOV of around 20.5 degrees, which is difficult to achieve with interferometer-based spectral imagers.

\section{Algorithm Design for Spectral Reconstruction}

\begin{figure}[ht!]
\centering\includegraphics[width=13cm]{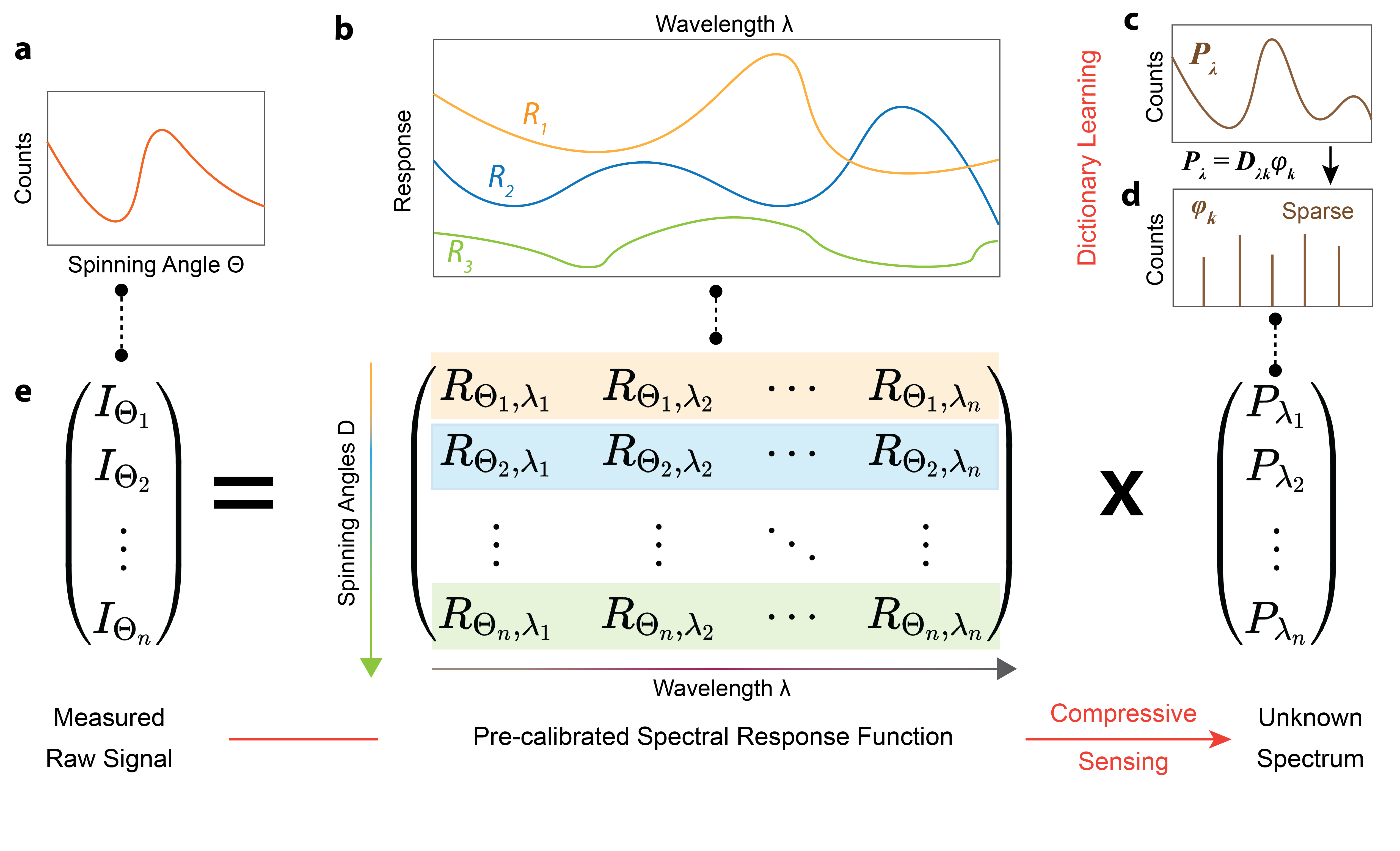}
\caption{Schematic of the spectral reconstruction process. The measured raw signal (a) can be expressed by the pre-calibrated spectral response function (b) of the imaging system multiplied by the spectrum of an imaging target (c). For the spectral reconstrcution, the unknown spectrum  $P_\lambda$ is projected onto a sparse representation basis $D_{k\lambda}$ using dictionary learning (c and d). This sparse representation $\phi_k$ is then used for compressive sensing based reconstruction (e). The use of compressive sensing and dictionary learning in the reconstruction process significantly improves the reconstruction accuracy, making the spinning-metasurface-based spectro-polarimetric imaging more robust against noise and measurement errors.}
\end{figure}

To extract the unknown spectro-polarimetric properties of various imaging targets, we use a combination of dictionary learning and compressive sensing algorithms in the reconstruction process. The tunable transmission spectra produced by our spinning metasurfaces (shown in Fig. 2f) are not narrowband, which means that the collected raw signals at different spinning-angle combinations $\Theta$ do not directly reflect the spectral radiance at different wavelengths. Instead, the collected signal $I(\Theta)$ at each pixel can be described as an integral of the spectral response function $R(\Theta, \lambda)$ multiplied by the ground truth spectrum $P(\lambda)$ that we wish to obtain, i.e. $I(\Theta)=\int_{\lambda_{\min }}^{\lambda_{\max }} R(\Theta, \lambda) P(\lambda) \mathrm{d} \lambda$. To solve for this equation, we discretize the spectral range of interest and express it in a tensor form as shown in Eq. \ref{eq:signal}:
\begin{equation}
I_{\Theta} = R_{\Theta\lambda}P_{\lambda}
\label{eq:signal}
\end{equation}

We emphasize that directly solving Eq.~\ref{eq:signal} does not produce accurate spectral reconstructions\cite{Planck, Mini2}. In theory, we can use measured signals $I_{\Theta}$ and the pre-calibrated response function $R_{\theta\lambda}$ to directly determine unknown spectra $P_\lambda$ at each pixel of a scene. However, in practice, two limitations impede the performance of spectral reconstruction: the problem becomes underdetermined when there are many discretized wavelength bands, and measurement noise affects both $I_{\Theta}$ and $R_{\Theta\lambda}$, making the direct reconstruction method unstable and the results inaccurate.

To enhance both the precision and consistency of spectral reconstruction, we leverage the capabilities of compressive sensing and dictionary learning algorithms to solve Eq.~\ref{eq:signal} (Fig.~3). Our approach begins with dictionary learning, a process we employ to generate a dictionary comprised of 32 basis functions, designated as $D$. The spectra utilized for the dictionary learning are derived  from the infrared emissivity spectra drawn from the ECOSTRESS Spectral Library \cite{NASA_Library1, NASA_Library2}, as well as blackbody thermal radiation spectra across various temperatures. The resultant dictionary can provide sparse representations for any thermal radiation spectrum in the space of spectra we are studying. We point out the efficiency of this sparse coding is fundamentally linked to the prevalent spectral similarities found within the LWIR thermal radiation spectra. We thus project the unknown spectrum $P_\lambda$ as a linear combination of the basis functions in the dictionary (Fig.~3 c and d). We have,
\begin{equation}
P_\lambda=D_{\lambda k} \phi_k
\end{equation}
where $\phi_k$ is a sparse coding of the spectrum $P_\lambda$. With this sparse representation, the spectral reconstruction problem can be solved by first obtaining $\phi_{\text{recon}}$:
\begin{equation}
\begin{gathered}
\phi_{\text {recon }}=\arg \min_{\phi_k}\|\phi_k\|_1 \\
\text { s.t. }\left\|I_\Theta-R_{\Theta \lambda} P_\lambda\right\|_2=\left\|I_\Theta-A_{\Theta k} \phi_k\right\|_2<\epsilon
\end{gathered}
\end{equation}
where $A_{\Theta k}=R_{\Theta\lambda}D_{\lambda k}$, and $\epsilon$ is the residual error. Finally, the spectra $P_{\lambda}$ at each pixel of a scene is reconstructed by 
\begin{equation}
P_{\text{recon}}=D_{_{\lambda k}}\phi_{\text{recon}}
\end{equation}

Our reconstruction method significantly improves the reconstruction accuracy, making the spinning-metasurface-based spectro-polarimetric imaging more robust against noise and measurement errors.

\section{Spectro-Polarimetric Imaging and Machine Vision}

\begin{figure}[ht!]
\centering\includegraphics[width=13cm]{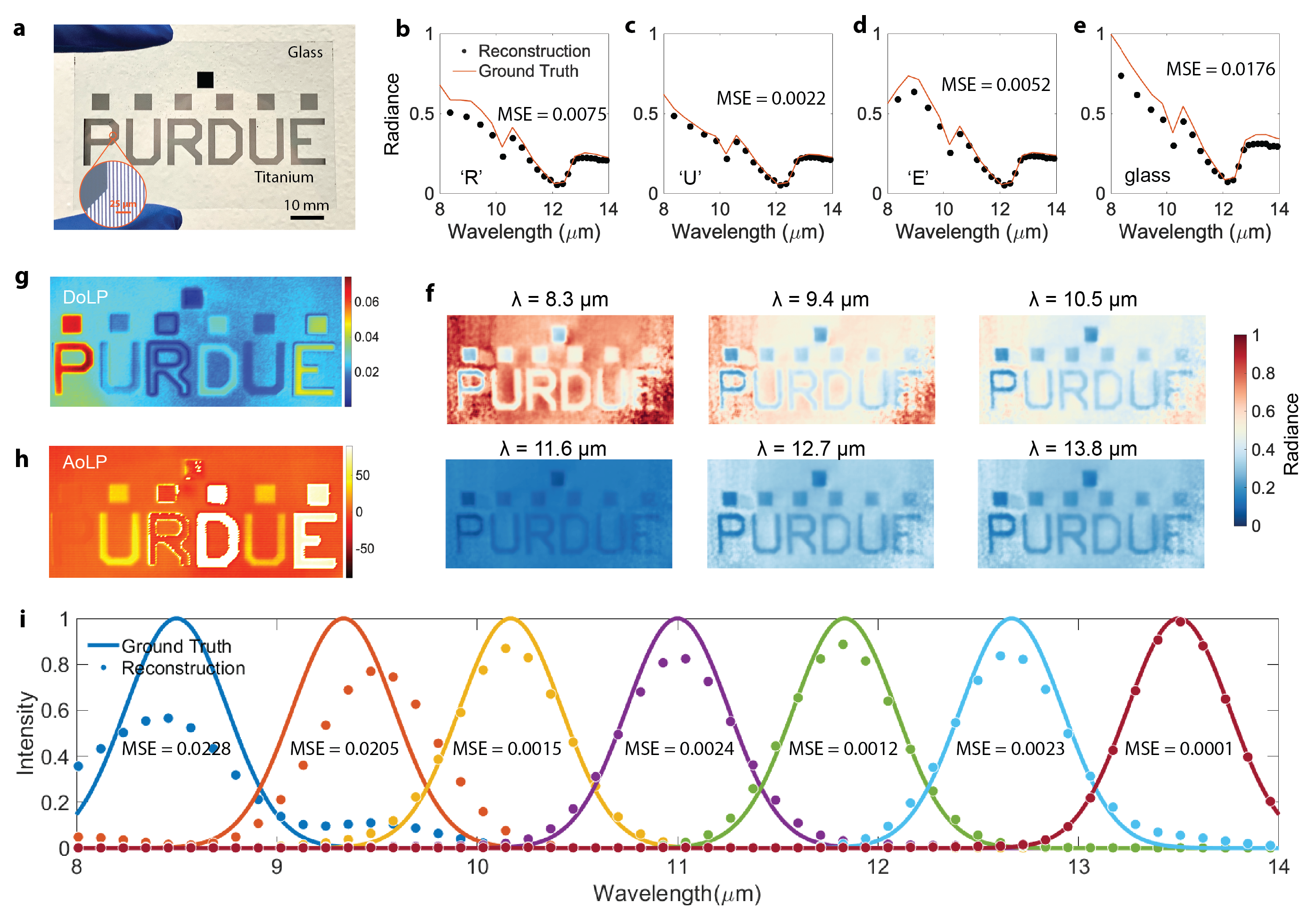}
\caption{Spectro-polarimetric thermal imaging results. (a) An optical image of the 'PURDUE' imaging target that is constructed from titanium letters on a glass substrate (75mm x 50mm). Inset: a zoomed-in optical microscope image of the micro-structures in the letters, which generate distinctive spectral and polarimetric signatures. (b-e) Reconstructed spectra of four representative pixels (corresponding to the letter 'R', 'U', 'E' and the glass substrate, respectively) compared with the ground truth spectra measured by a Fourier-transform infrared spectrometer. (f) Reconstructed spectral frames at 6 representative wavelengths. The contrast between different frames demonstrates that the system can effectively reveal the LWIR spectral properties of various materials and structures. (g-h) Degree-of-linear-polarization and angle-of-linear-polarization frames. Distinctive polarimetric signatures can be observed for each letter in the images. (i) Simulated spectral reconstruction results. The ground truth spectra (solid lines) are Gaussian peaks with 0.6~$\mu m$  FWHM centered at different wavelengths (8.5~$\mu m$, 9.3~$\mu m$, 10.2~$\mu m$, 11~$\mu m$, 11.8~$\mu m$, 12.7~$\mu m$, and 13.5~$\mu m$). The reconstructed spectra (dotted lines) show good agreements with the ground truth.}
\end{figure}

To evaluate the performance of our prototype imaging system, we conduct experiments using a custom-designed "PURDUE" target made of letters constructed from titanium and a glass substrate (Fig.~4a). Each letter has unique micro-grating structures (Inset of Fig.~4a) that generate distinctive spectro-polarimetric signatures in the thermal radiation signal. The glass substrate also features a characteristic emission peak around 11 $\mu m$. Note that we heat the image target to 150$^\circ C$ to generate high signal intensity. The reconstructed spectra of four representative pixels are shown in Fig.~4 b-e. We compare them with the ground truth spectra measured by a Fourier-transform infrared spectrometer, validating the effectiveness of our reconstruction approach. The reconstructed spectral frames (Fig.~4f) also exhibit high contrasts between different wavelengths, demonstrating that the system can effectively reveal the LWIR spectral properties of different targets. We note that the relatively low reconstruction accuracy at shorter wavelengths (8 - 10 $\mu m$) results from the low transmission (low SNR) and the high correlation (similarity) between the tuned spectra (Fig. 2f).

To characterize the spectral resolution of our system, we conduct a numerical simulation of the spectral reconstruction performance using Gaussian peaks centered at varying wavelengths. This testing method is a widely adopted way to determine the fundamental resolution constraints of reconstruction-based spectroscopy \cite{MosaicSensor0, MosaicSensor1, DiffuserCam, Mini2, Mini1, Planck}. The simulation incorporates the tunable transmission spectrum set (as depicted in Fig.~2f) produced by the spinning-metasurface module, and utilizes the same dictionary learning and compressive sensing algorithms previously discussed. Fig.~4i reveals that Gaussian peaks with a 0.6 µm full width at half maximum (FWHM) centered above 10 µm can be precisely reconstructed with negligible mean square errors (MSE). Additional simulations also show that our system is capable of detecting narrower spectral peaks with FWHM as low as 0.1 micron (see Supplemental Document figure S4 for details). We again observe that the performance of spectral reconstruction is dependent on wavelength. A higher level of accuracy is achieved at longer wavelengths due to the significant differences between the tuned transmission spectra in this region (Fig.~2f). This implies that we could further enhance the spectral resolving power by refining the design of the metasurfaces and generate a broader array of mutually uncorrelated tunable transmission spectra.

We also obtain the polarimetric information including degree of the linear polarization (DOLP) and the angle of linear polarization (AoLP) using the designed system. For polarimetric imaging, we collectively rotate the spinning metasurfaces and the input polarizer, selecting four different polarizations ($0^{\circ}$, $90^{\circ}$, $45^{\circ}$ and $-45^{\circ}$) while maintaining the same spectral transmission. We use the first three Stokes parameters to quantify the polarimetric information associated with each pixel, i.e. $S_{0} =I_{0}+I_{90}$, $S_{1} =I_{0}-I_{90}$, and $S_{2} =I_{45}-I_{-45}$, where $I_{0}$, $I_{90}$, $I_{45}$ and $I_{-45}$ are the light intensity at polarization angles of $0^{\circ}$, $90^{\circ}$, $45^{\circ}$ and $-45^{\circ}$, respectively. The DoLP and AoLP are then calculated at each wavelength through $\mathrm{DoLP}=\sqrt{S_{1}^{2}+S_{2}^{2}} / S_{0}$ and $\mathrm{AoLP}=arctan (S_{2}/S_{1})$. As shown in Fig.~4 g and h, we can clearly distinguish between different letters based on their polarimetric signatures in the thermal radiation signal. The four-dimensional spatial-spectro-polarimetric data-tesseract provides significantly more insight associated with an object, making it a powerful tool for a wide range of imaging applications. 

\begin{figure}[ht!]
\centering\includegraphics[width=13cm]{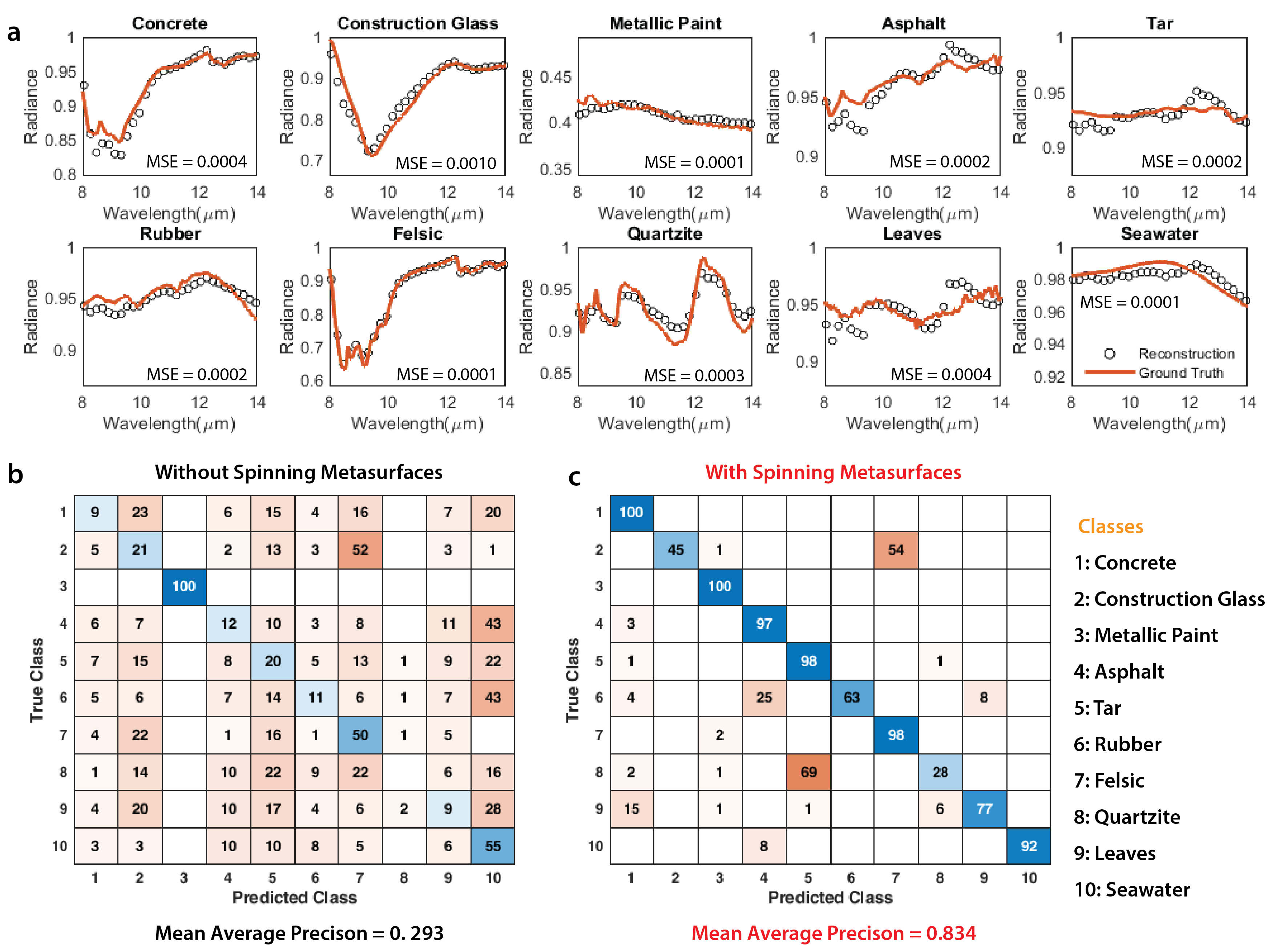}
\caption{Material classification using Spinning MetaCam. (a) Simulated spectral reconstruction results for materials commonly found in everyday life. (b) Calculated confusion matrix illustrating material classification without the use of spinning metasurfaces. The classification in this case is solely reliant on the integrated intensity of thermal radiation within the LWIR region. (c) Calculated confusion matrix presenting material classification with spinning metasurfaces incorporated. The inclusion of spectral information substantially augments the performance of material classification, improving the mean average precision from 0.293 to 0.834.}
\end{figure}

To prove the effectiveness and reveal the potential of Spinning MetaCam, we conduct numerical tests demonstrating its applicability in material classification. Fig.~5a shows that our spinning metasurfaces can accurately reconstruct the LWIR thermal radiation spectra of a diverse range of materials commonly found in everyday life (ground truth obtained from the ECOSTRESS Spectral Library \cite{NASA_Library1, NASA_Library2}). These  results also highlight the spectral fidelity and universality of our spectral reconstruction method. Employing these thermal radiation spectra, we execute material classification by identifying the material whose ground truth spectrum exhibits the least mean square error (MSE) when compared to the reconstructed spectrum. We repeat this spectral reconstruction and material detection procedure 100 times for each material, and use the outcomes to generate a confusion matrix (see Supplemental Document for details). Fig.~5b and Fig.~5c compare the resulting confusion matrices derived with and without the use of spinning metasurfaces, while maintaining a consistent noise level of 5\%. Without the spinning metasurfaces, material classification relies solely on the integrated intensity of the thermal radiation over the LWIR region. This approach exhibits a low precision for most materials, being effective only for those materials with substantially disparate thermal radiation intensities, such as metallic paint. In stark contrast, material classification significantly improves in accuracy when the spectra collected by the Spinning MetaCam is utilized. We note that instances of lower accuracy only arise when there is a pronounced similarity in the LWIR spectra of materials, as observed between construction glass and felsic. For a more quantitative analysis, we perform the material classification test at varying noise levels ranging from 1\% to 10\%, and subsequently calculate the mean average precision (mAP)\cite{Yolo}. Representing the overall accuracy of classification for all classes of materials at all different noise levels, the mAP reaches an impressively high value of 0.834, providing a significant improvement from the low mAP of 0.293 obtained without the spinning metasurfaces.

\section{Conclusion}
Our results provide an innovative approach for spectro-polarimetric thermal imaging by combining meta-optics and computational spectral reconstruction. The low-SWaP (size, weight, and power) system opens the door for physics-driven machine vision. The high-dimensional thermal image data can significantly improve the performance of tasks such as depth estimation, object detection, and semantic segmentation when only radiative heat signal is available. Furthermore, we foresee that spectro-polarimetric thermal imaging can also be a powerful tool for scientific research, allowing for non-destructive characterization in the infrared region to investigate a wide range of novel physical phenomena, such as anisotropic thermal conduction\cite{Anisotropy} and directional or nonreciprocal radiative heat transfer \cite{Direction, yang2022polarimetric, shayegan2023direct}. Overall, our work provides a key development in the rapidly growing field of thermal imaging, offering a pivotal technology for heat-assisted detection and ranging.

\begin{backmatter}
\bmsection{Funding}
This work was supported by the United States Department of Energy, Office of Basic Sciences under DE-SC0017717, and the Defence Advanced Research Projects Agency (DARPA) under  the  Nascent  Light-Matter Interactions (NLM) program.


\bmsection{Disclosures}
The authors declare no conflicts of interest.

\bmsection{Data availability} Data underlying the results presented in this paper are not publicly available at this time but may be obtained from the authors upon reasonable request.

\bmsection{Supplemental document}
See Supplemental Document for supporting content. 

\end{backmatter}

\section{References}

\bibliography{sample}

\end{document}